\documentclass[symmetry,article,accept,pdftex,moreauthors]{Definitions/mdpi}
\usepackage{amsmath}
\usepackage{amsfonts}
\usepackage{amssymb,bm}
\usepackage{siunitx}
\usepackage{color}
\usepackage{braket}
\usepackage{graphics}

\firstpage{1}
\pubvolume{15}
\issuenum{8}
\articlenumber{1580}
\pubyear{2023}
\copyrightyear{2023}
\datereceived{5 July 2023}
\daterevised{8 August 2023} 
\dateaccepted{9 August 2023}
\datepublished{13 August 2023 }
\hreflink{https://doi.org/10.3390/sym15081580} 

\Title{Nonequilibrium Casimir-Polder Interaction Between Nanoparticles and Substrates
Coated with Gapped Graphene}

\TitleCitation{Nonequilibrium Casimir-Polder Interaction Between Nanoparticles and Substrates
Coated with Gapped Graphene}

\Author{{Galina L. Klimchitskaya} ${}^{1,2}$\orcidA{},
Constantine C. Korikov${}^{3}$,
Vladimir M. Mostepanenko ${}^{1,2,4,}$*\orcidB{} and
Oleg {Yu.} Tsybin ${}^{2}$}

\AuthorNames{Galina L. Klimchitskaya, Constantine C. Korikov,
Vladimir M. Mostepanenko, Oleg Yu. Tsybin}

\AuthorCitation{Klimchitskaya, G.L.; Korikov, C.C.; Mostepanenko, V.M.; Tsybin, O.Y.}

\address{%
$^{1}$ \quad Central Astronomical Observatory at Pulkovo of the Russian Academy of Sciences, 196140 Saint Petersburg, Russia\\
$^{2}$ \quad Peter the Great Saint Petersburg
Polytechnic University, 195251 Saint Petersburg, Russia\\
$^{3}$ \quad Huawei Noah's Ark Lab, Krylatskaya str. 17, Moscow 121614, Russia \\
$^{4}$ \quad {Kazan Federal University}, 420008 Kazan, Russia}

\corres{Correspondence: vmostepa@gmail.com}

\abstract{The out-of-thermal-equilibrium Casimir-Polder force between
nanoparticles and dielectric substrates coated with gapped graphene is
considered in the framework of the Dirac model using the formalism of the
polarization tensor. This is an example of physical phenomena violating
the time-reversal symmetry. After presenting the main points of the used
formalism, we calculate two contributions to the Casimir-Polder force
acting on a nanoparticle on the source side of a fused silica glass
substrate coated with gapped graphene, which is either cooler or hotter
than the environment. The total nonequilibrium force magnitudes are
computed as a function of separation for different values of the energy
gap and compared with those from an uncoated plate and with the equilibrium
force in the presence of graphene coating. According to our results, the
presence of a substrate increases the magnitude of the nonequlibrium force.
The force magnitude becomes larger with higher and smaller with lower
temperature of the graphene-coated substrate as compared to the equilibrium
force at the environmental temperature. It is shown that with increasing
energy gap the magnitude of the nonequilibrium force becomes smaller, and
the graphene coating makes a lesser impact on the force acting on a
nanoparticle from the uncoated substrate. Possible applications of the
obtained results are discussed.}


\begin{document}


\section{Introduction}

Much use is made of the physical systems which parts are kept at different
temperatures, i.e., in out-of-thermal-equilibrium conditions. The
processes occurring in thermally nonequilibrium systems are usually
irreversable, i.e., they lack of symmetry inherent in systems characterized
by some unique temperature equal to that of the environment. The theoretical
description of thermally nonequilibrium systems is a rather complicated
subject. However, the demands of both fundamental physics and its
applications in nanotechnology lent impetus to a successful search for new
theoretical approaches.

The Casimir-Polder interaction is an example on how the standard theory
of a thermally equilibrium phenomenon can be extended to the
out-of-thermal-equilibrium conditions. Casimir and Polder \cite{1} derived
an expression for the force acting between a small polarizable particle
and an ideal metal plane kept at zero temperature. In the framework of
the Lifshitz theory \cite{2}, this result was generalized to the case
when an ideal metal plane is replaced with thick material plate kept
at the same temperature $T_E$ as the environment \cite{3,4}. The
Casimir-Polder force is a thermally equilibrium quantum phenomenon
determined by the zero-point and thermal fluctuations of the
electromagnetic field. It finds extensive applications in atomic physics
and condensed matter physics (see the monographs \cite{5,6,7,8} for a
large body of examples). Specifically, much attention is given to the
Casimir-Polder interaction between nanoparticles and material surfaces
of different nature including biomembranes
\cite{9,10,11,12,13,14,15,16,17,18}.

The extension of the Lifshitz theory to physical systems out of thermal
equilibrium, which covers both cases of two macroscopic bodies and a small
particle near the surface of a macroscopic body, was developed in
\cite{19,20,21,22,23,24} and further elaborated in
\cite{25,26,27,28,29,30,31,32}. This was done under an assumption that each
body is in the state of local thermal equilibrium \cite{22}. The developed
theory was confirmed experimentally by measuring the Casimir-Polder force
between $^{87}$Rb atoms and a fused silica glass plate heated up to the
temperatures much higher than $T_E$ \cite{33}. The case when the dielectric
properties of a plate depend heavily on the temperature offers a fertile
field for the study of the nonequilibrium Casimir-Polder force. This is
true the for metallic plates \cite{28,30} and for the plates made of a
material, which undergoes the phase transition with increasing temperature
\cite{29}.

The material, which demonstrates a profound effect of temperature on its
dielectric properties, is graphene, i.e., a one-atom-thick layer of carbon
atoms packed in the hexagonal lattice \cite{34}. At low energies, graphene
is well-described by the Dirac model \cite{35,36}. Due to its relatively
simple structure, the spatially nonlocal dielectric properties of graphene
can be expressed via the polarization tensor in (2+1) dimensions
\cite{37,38} and found on the solid basis of quantum electrodynamics at any
temperature \cite{39,40,41,42}. This presents the immediate possibility
to calculate the Casimir-Polder force between nanoparticles and either
heated or cooled graphene sheets using the extension of the Lifshits
theory to the out-of-thermal-equilibrium situations. Calculations of
this kind are of prime interest not only for fundamental physics, but for
numerous applications harnessing interaction of nanoparticles with
graphene and graphene-coated substrates as well (see, e.g.,
\cite{43,44,45,46,47,48}).

The investigation of the nonequilibrium Casimir-Polder interaction between
nanoparticles and graphene originated in \cite{49}, where the freestanding
in vacuum pristine graphene sheet was considered. The pristine character of
graphene means that there is no energy gap in the spectrum of quasiparticles,
i.e., they are massless, and that it possesses the perfect crystal lattice
with no foreign atoms. Both these assumptions are in the basis of the
original Dirac model of graphene \cite{34,35,36}. According to \cite{49},
the nanoparticles have the same temperature as the environment, whereas
the graphene sheet may be either cooler or hotter than the environment,
which takes the nanoparticle-graphene system out of the state of thermal
equilibrium. It was shown \cite{49} that the nonequilibrium conditions
have strong effect on the nanoparticle-graphene force and can even
change the Casimir-Polder attraction with repulsion if the temperature of
a graphene sheet is lower than $T_E$.

Taking into account that real graphene sheets are characterized by some
energy gap, i.e., small but nonzero mass of quasiparticles \cite{34},
an impact of the energy gap on the nonequilibrium Casimir-Polder force
between a nanoparticle and a freestanding gapped graphene sheet was
investigated in \cite{50}. It was shown that by varying the energy gap
it is possible to control the force value. Unlike the case of a pristine
graphene, the force acting on a nanoparticle from the sheet of gapped
graphene preserves its attractive character.

When employing graphene sheets in physical experiments or in micro- and
nanodevices, they are usually not freestanding in vacuum, but deposited
on some dielectric substrate. Because of this, it is important to
determine an impact of substrate underlying the gapped graphene sheet on
the nonequilibrium Casimir-Polder force acting on a nanoparticle. In this
article, we investigate the force on a spherical nanoparticle from the
source side of dielectric substrate coated with gapped graphene, which
is either cooled or heated as compared to the environmental temperature
$T_E$. As to the temperature of nanoparticles, it is assumed to be equal
to $T_E$.

After presenting the main points of the theoretical formalism, we continue
with a comparison between the equilibrium Casimir-Polder forces on a
nanoparticle from the uncoated and coated with gapped graphene dielectric
substrate. Then, the nonequilibrium Casimir-Polder force acting on a
nanoparticle from an uncoated substrate is considered. Next, we turn our
attention to the calculation of two contributions of different nature to
the nonequilibrium force on a nanoparticle from the graphene-coated
substrates with various values of the energy gap of graphene coating kept
at different temperatures. Finally, the total nonequilibrium force on a
nanoparticle is studied as the function of separation at different
temperatures of a graphene-coated substrate and typical values of the
energy gap.

It is shown that the presence of a substrate increases the magnitude
of the nonequilibrium force, which is larger and smaller than the
equilibrium one for the graphene-substrate temperature higher and
lower than $T_E$, respectively. In all cases, an increase of the
energy gap leads to smaller force magnitudes and to a lesser impact
of the graphene coating on the nonequilibrium force on a nanoparticle
from the uncoated dielectric substrate.

The article is organized as follows. Section 2 presents the necessary
information from the theory of nonequilibrium Casimir-Polder force
acting on nanoparticles from the substrates coated with gapped graphene.
In Section 3, both the equilibrium and nonequilibrium forces on
nanoparticles from the fused silica glass plate are considered and
compared with the equilibrium one in the presence of graphene coating.
In Section 4, the nonequilibrium Casimir-Polder force from a fused
silica plate coated with gapped graphene is investigated. Section 5
contains the discussion. In Section 6, the reader will find our
conclusions.

\section{Theoretical Description of Nonequilibrium Casimir-Polder Interaction from
Graphene-Coated Substrates}
\newcommand{\adt}{(a,\Delta,T_E,T_g)}
\newcommand{\ok}{(\omega,k,\Delta,T_g)}
\newcommand{\wk}{(\omega,k)}
\newcommand{\fok}{(\omega,k,\Delta)}
\newcommand{\sok}{(\omega,k,\Delta)}
\newcommand{\tp}{\tilde{p}}
\newcommand{\Fv}{{\tilde{v}_F}}
\newcommand{\Sv}{{\tilde{v}_F^2}}
\newcommand{\ve}{{\varepsilon}}
\newcommand{\Mr}{{R_{\rm TM}^{\rm sub}}}
\newcommand{\Er}{{R_{\rm TE}^{\rm sub}}}
\newcommand{\oT}{{(\omega,k,\Delta,T_g)}}
\newcommand{\qt}{{\tilde{q}}}
\newcommand{\dt}{{\widetilde{\Delta}}}
\newcommand{\tP}{{\widetilde{\Pi}}}
\newcommand{\zgg}{{(t,\widetilde{\Delta})}}
\newcommand{\ogg}{{(t,\widetilde{\Delta},\tau_g)}}
\newcommand{\yFt}{{\sqrt{1-\tilde{v}_F^2t^2}}}

Here, we briefly present the formalism required for calculation of the Casimir-Polder
force between nanoparticles and dielectric substrates coated with gapped graphene in
out-of-thermal-equilibrium conditions. We assume that nanoparticles and the environment have
the temperature $T_E$ (in computations we use $T_E=300~$K), whereas the graphene-coated
substrate has the temperature $T_g$, which may be either lower or higher than $T_E$.
The separation distance between nanoparticles and the plane surface of area $S$ of
a graphene-coated substrate is $a$. The energy gap of a graphene coating is denoted
$\Delta$. Below we consider the simplest case of spherical nanoparticles whose radius
$R$ satisfies the conditions

\begin{equation}
R\ll a\ll\frac{\hbar c}{k_BT_{E,g}},
\label{eq1}
\end{equation}

\noindent
where $k_B$ is the Boltzmann constant. It is suggested also that $a\ll\sqrt{S}$.

Nanoparticles of arbitrary shape spaced at any separation from the surface, which
should not be necessarily plane, can be considered using the more complicated
scattering matrix approach \cite{24,27}.

It is convenient to represent the nonequilibrium Casimir-Polder force as a sum of two
contributions \cite{22,24}

\begin{equation}
F_{\rm neq}\adt =F_{M}\adt+F_r\adt,
\label{eq2}
\end{equation}

\noindent
Here, $F_M$ somewhat resembles the equilibrium Casimir-Polder force expressed as a sum over
the discrete Matsubara frequencies, whereas $F_r$ is the proper nonequilibrium contribution.

Actually, both terms in (\ref{eq2}) describe some part of the effects of nonequilibrium and
their explicit forms depend on which of the items responsible for these effects are
included in $F_M$ and which in $F_r$. Below we employ $F_M$ and $F_r$ in the forms derived
in \cite{49} and used in \cite{50}.  Thus, $F_M$ has the form

\begin{eqnarray}
&&
F_{M}\adt=-\frac{2k_BT_E\alpha(0)}{c^2}\sum_{l=0}^{\infty}{\vphantom{\sum}}^{\prime}
\int\limits_0^{\infty}k\,dke^{-2aq_l(k)}
\label{eq3}\\
&&~~~
\times
\left\{\left[2q_l^2(k)c^2-\xi_{E,l}^2\right]
\Mr(i\xi_{E,l},k,\Delta,T_g)-\xi_{E,l}^2\Er(i\xi_{E,l},k,\Delta,T_g)\right\},
\nonumber
\end{eqnarray}

\noindent
where $\alpha(0)$ is the static polarizability of a nanoparticle, $\mbox{\boldmath$k$}$ is
the component of the wave vector parallel to the plane of graphene,
$k=|\mbox{\boldmath$k$}|$, the prime on the sum divides the term with $l=0$ by 2, and
the Matsubara frequencies and $q_l$ are defined as

\begin{equation}
\xi_{E,l}=\frac{2\pi k_BT_El}{\hbar},\qquad
q_l^2(k)=k^2+\frac{\xi_{E,l}^2}{c^2}.
\label{eq4}
\end{equation}

The quantities $R_{\rm TM,TE}^{\rm sub}$ are the reflection coefficients of the electromagnetic
waves on the graphene-coated substrate for the transverse magnetic and transverse electric
polarizations. In fact (\ref{eq3}) has the same form as the Lifshitz formula for the
equilibrium Casimir-Polder force between a nanoparticle and a plate, but with one important
difference. Here, the Matsubara frequencies are defined at the environmental temperature
$T_E$, but the reflection coefficients --- at the temperature of graphene-coated substrate
$T_g$ (in the Lifshitz formula $T_E=T_g$). Note also that (\ref{eq3}) contains the static
polarizability $\alpha(0)$ in front of the summation sign in $l$, whereas in the Lifshitz
formula for an atom-wall interaction $\alpha(i\xi_l)$ appears under the sum in $l$.
This is because the dynamic polarizability of nanoparticles at several first Matsubara
frequencies contributing to the force under the condition (\ref{eq1}) reduces to
$\alpha(i\xi_l)\approx\alpha(0)$ \cite{26}.

The proper nonequilibrium contribution $F_r$ in (\ref{eq2}) is given by \cite{49,50}

\begin{eqnarray}
&&
F_{r}\adt=\frac{2\hbar\alpha(0)}{\pi c^2}
\int\limits_{0}^{\infty}\!\!d\omega\,\Theta(\omega,T_E,T_g)\!\!
\int\limits_{\omega/c}^{\infty}\!\!k\,dke^{-2aq(\omega,k)}
\nonumber \\
&&~~~~
\times
{\rm Im}\left\{\left[2q^2\wk c^2+\omega^2\right]
\Mr\ok+\omega^2\Er\ok\right\}.
\label{eq5}
\end{eqnarray}

\noindent
where
\begin{equation}
\Theta(\omega,T_E,T_g)=\frac{1}{\exp\left(\frac{\hbar\omega}{k_BT_E}\right)-1}-
\frac{1}{\exp\left(\frac{\hbar\omega}{k_BT_g}\right)-1}, \qquad
q^2\wk=k^2-\frac{\omega^2}{c^2}.
\label{eq6}
\end{equation}

\noindent
It is presented as an integral over the real frequency axis. In so doing, only
$k>\omega/c$, i.e., only the evanescent waves, contribute to (\ref{eq5}).

In the literature, the other forms of $F_M$ and $F_r$ have been considered, where
$F_r$ contains contributions from both the evanescent and propagating waves \cite{22}.
The advantage of (\ref{eq5}) is a presence of the exponential factor with real $q$ and
negative power which secures the quick convergence of the integral.

The reflection coefficients on the graphene-coated substrate were expressed via
the frequency-dependent dielectric permittivity of substrate $\ve(\omega)$ and
the polarization tensor of graphene $\Pi_{mn}\ok$ in (2+1) dimensions, i.e., with
$m,\,n=0,\,1,\,2$ \cite{51}

\begin{eqnarray}
&&
\Mr\ok=\frac{\hbar k^2[\ve(\omega)q\wk-q_{\ve}\wk]+q\wk q_{\ve}\wk
\Pi_{00}\ok}{\hbar k^2[\ve(\omega)q\wk+q_{\ve}\wk]+q\wk q_{\ve}\wk
\Pi_{00}\ok},
\nonumber\\[-1mm]
&&\label{eq7}\\[-1mm]
&&
\Er\ok=\frac{\hbar k^2[q\wk-q_{\ve}\wk]-\Pi\ok}{\hbar k^2[q\wk+q_{\ve}\wk]+\Pi\ok},
\nonumber
\end{eqnarray}

\noindent
where
\begin{eqnarray}
&&
q_{\ve}^2\wk=k^2-\ve(\omega)\frac{\omega^2}{c^2},
\label{eq8}\\
&&
\Pi\ok=k^2\Pi_{m}^{\,m}\ok-q^2\wk\Pi_{00}\ok.
\nonumber
\end{eqnarray}

For an uncoated substrate $\Pi_{00}=\Pi=0$ and (\ref{eq7}) transforms into the
familiar Fresnel reflection coefficients. To obtain the reflection coefficients at the pure
imaginary frequencies entering (\ref{eq3}), one should put in (\ref{eq7})
$\omega=i\xi_{E,l}$. Here, we also assume that the dielectric permittivity of
a substrate material does not depend on temperature, contrary to the polarization
tensor of graphene. In the case of materials (metals, for instance)
with the temperature-dependent dielectric
permittivity, one should change $\ve(\omega)$ in (\ref{eq7}) with $\ve(\omega,T_g)$.

Computations of the nonequilibrium Casimir-Polder force between nanoparticles and
graphene-coated substrate can be performed by Equations (\ref{eq2}), (\ref{eq3}),
(\ref{eq5}), and (\ref{eq7}) if one knows the explicit expressions for the components
of the polarization tensor of graphene. For the gapped graphene, these components
defined along both real and imaginary frequency axes were found in \cite{41} and
presented more specifically in \cite{49,50} in terms of the variables $\omega$ and $k$.
It should be noted, however, that the numerical computations of (\ref{eq5}) along
the real frequency axis for substrates coated with gapped graphene are much more
involved than of (\ref{eq3}) and should be performed using the convenient choice
of dimensionless variables. That is why, below we introduce the necessary
dimensionless quantities and present the results (\ref{eq5})--(\ref{eq8}) and the
expressions for the dimensionless polarization tensor in these terms.

Let us define the following dimensionless quantities

\begin{eqnarray}
&&
t=\frac{ck}{\omega},\qquad \qt(t)=\sqrt{t^2-1}=\frac{c}{\omega}q\wk,
\qquad \qt_{\ve}(\omega,t)=\sqrt{t^2-\ve(\omega)}=\frac{c}{\omega}q_{\ve}\wk,
\nonumber \\[-1mm]
&&\label{eq9} \\[-1mm]
&&
\tP_{00}\ogg=\frac{c}{\hbar\omega t^2}\Pi_{00}\ok, \qquad
\tP\ogg=\frac{c^3}{\hbar\omega^3 t^2}\Pi\ok,
\nonumber
\end{eqnarray}

\noindent
where

\begin{equation}
\dt=\frac{\Delta}{\hbar\omega}, \qquad \tau_g=\frac{\hbar\omega}{k_BT_g}.
\label{eq10}
\end{equation}

Using (\ref{eq9}) and (\ref{eq10}), the reflection coefficients (\ref{eq7})
can be rewritten as

\begin{eqnarray}
&&
\Mr(\omega_cx,t,\dt,\tau_g)=\frac{\ve(\omega_cx)\qt(t)-\qt_{\ve}(\omega_cx,t)+
\qt(t)\qt_{\ve}(\omega_cx,t)\tP_{00}\ogg}{\ve(\omega_cx)\qt(t)+\qt_{\ve}(\omega_cx,t)+
\qt(t)\qt_{\ve}(\omega_cx,t)\tP_{00}\ogg},
\nonumber \\[-1mm]
&& \label{eq11} \\[-1mm]
&&
\Er(\omega_cx,t,\dt,\tau_g)=\frac{\qt(t)-\qt_{\ve}(\omega_cx,t)-
\tP\ogg}{\qt(t)+\qt_{\ve}(\omega_cx,t)+\tP\ogg},
\nonumber
\end{eqnarray}

\noindent
where

\begin{equation}
x=\frac{\omega}{\omega_c(a)}, \qquad \omega_c(a)=\frac{c}{2a}.
\label{eq12}
\end{equation}

Below in this section, we rewrite the results of \cite{41,50} for the polarization tensor
in the region of evanescent waves in terms of dimensionless variables (\ref{eq9}) and
(\ref{eq10}). It is common to present the polarization tensor as the sum of two
contributions

\begin{eqnarray}
&&
\tP_{00}\ogg=\tP_{00}^{(0)}\zgg+\tP_{00}^{(1)}\ogg,
\nonumber \\
&&
\tP\ogg=\tP^{(0)}\zgg+\tP^{(1)}\ogg.
\label{eq13}
\end{eqnarray}

The first terms on the right-hand side of (\ref{eq13}) refer to the case of zero
temperature, whereas the second have the meaning of the thermal corrections to them.
The polarization tensor in the area of evanescent waves $k>\omega/c$ has different
forms in the so-called plasmonic region \cite{52}

\begin{equation}
\frac{\omega}{c}<k<\frac{\omega}{v_F}, \qquad 1<t<\frac{1}{\Fv}\approx 300
\label{eq14}
\end{equation}

\noindent
and in the region

\begin{equation}
k>\frac{\omega}{v_F}, \qquad t>\frac{1}{\Fv}\approx 300,
\label{eq15}
\end{equation}

\noindent
where $v_F\approx c/300$ is the Fermi velocity for graphene and $\Fv=v_F/c$.

We begin with the plasmonic region (\ref{eq14}). Here, the polarization tensor is also
defined differently depending on the fulfilment of some condition. Thus, if the following
inequality is satisfied:

\begin{equation}
\frac{\dt}{\yFt}>1,
\label{eq16}
\end{equation}

\noindent
the first contributions to the polarization tensor (\ref{eq13}) are given by

\begin{eqnarray}
&&
\tP_{00}^{(0)}\zgg=-\frac{2\alpha}{\yFt}\,\Phi_A\left(\frac{\dt}{\yFt}\right),
\nonumber\\[-1mm]
&&\label{eq17}\\[-1mm]
&&
\tP^{(0)}\zgg={2\alpha}{\yFt}\,\Phi_A\left(\frac{\dt}{\yFt}\right),
\nonumber
\end{eqnarray}

\noindent
where $\alpha=e^2/(\hbar c)$ is the fine structure constant and

\begin{equation}
\Phi_A(x)=x-(1+x^2)\,{\rm arctanh}\frac{1}{x}.
\label{eq18}
\end{equation}

Under the condition (\ref{eq16}), the second contributions to (\ref{eq13}) take the form

\begin{adjustwidth}{-\extralength}{0cm}
\begin{eqnarray}
&&
\tP_{00}^{(1)}\ogg=\frac{8\alpha}{\Sv t^2}\int\limits_{\dt}^{\infty}dz\Psi(z,\tau_g)
\left[1-\frac{1}{2\yFt}\sum_{\lambda=\pm 1}\lambda
\frac{(z+\lambda)^2-\Sv t^2}{\sqrt{(z+\lambda)^2-\Sv t^2A(\dt,t)}}\right],
\nonumber\\[-0.5mm]
&&\label{eq19} \\[0.5mm]
&&
\tP^{(1)}\ogg=\frac{8\alpha}{\Sv t^2}\int\limits_{\dt}^{\infty}dz\Psi(z,\tau_g)
\left[1-\frac{1}{2}\yFt\sum_{\lambda=\pm 1}\lambda
\frac{(z+\lambda)^2-\Sv t^2[1-A(\dt,t)]}{\sqrt{(z+\lambda)^2-\Sv t^2A(\dt,t)}}\right],
\nonumber
\end{eqnarray}
\end{adjustwidth}

\noindent
where

\begin{equation}
A(\dt,t)=1-\frac{\dt^2}{1-\Sv t^2}, \qquad \Psi(z,\tau_g)=\frac{1}{e^{\tau_gz}+1}.
\label{eq20}
\end{equation}

{}From (\ref{eq17}) and (\ref{eq19}), it is seen that under the inequality (\ref{eq16})
the polarization tensor is real.

Next, we continue the consideration of the plasmonic region (\ref{eq14}) but under the
inequality opposite to (\ref{eq16})

\begin{equation}
\frac{\dt}{\yFt}<1.
\label{eq21}
\end{equation}

\noindent
In this case, the first contributions to the polarization tensor (\ref{eq13})
take the form

\begin{eqnarray}
&&
\tP_{00}^{(0)}\zgg=-\frac{2\alpha}{\yFt}\,\Phi_B\left(\frac{\dt}{\yFt}\right),
\nonumber\\[-1mm]
&&\label{eq22}\\[-1mm]
&&
\tP^{(0)}\zgg={2\alpha}{\yFt}\,\Phi_B\left(\frac{\dt}{\yFt}\right),
\nonumber
\end{eqnarray}

\noindent
where

\begin{equation}
\Phi_B(x)=x-(1+x^2)\left({\rm arctanh}\,{x}+i\frac{\pi}{2}\right).
\label{eq23}
\end{equation}

The second contributions to (\ref{eq13}) under the inequality (\ref{eq21}) take
a more complicated form

\begin{adjustwidth}{-\extralength}{0cm}
\begin{eqnarray}
&&
\tP_{00}^{(1)}\ogg=\frac{8\alpha}{\Sv t^2}\left\{\int\limits_{\dt}^{z^{(-)}}dz\Psi(z,\tau_g)
\left[1-\frac{1}{2\yFt}\sum_{\lambda=\pm 1}
\frac{(z+\lambda)^2-\Sv t^2}{\sqrt{(z+\lambda)^2-\Sv t^2A(\dt,t)}}\right]\right.
\nonumber \\[0.5mm]
&&
~~~~~~~~
+\left.\int\limits_{z^{(-)}}^{\infty}dz\Psi(z,\tau_g)
\left[1-\frac{1}{2\yFt}\sum_{\lambda=\pm 1}\lambda
\frac{(z+\lambda)^2-\Sv t^2}{\sqrt{(z+\lambda)^2-\Sv t^2A(\dt,t)}}\right]\right\},
\nonumber\\[-0.5mm]
&&\label{eq24} \\[0.5mm]
&&
\tP^{(1)}\ogg=\frac{8\alpha}{\Sv t^2}\left\{\int\limits_{\dt}^{z^{(-)}}dz\Psi(z,\tau_g)
\left[1-\frac{1}{2}\yFt\sum_{\lambda=\pm 1}
\frac{(z+\lambda)^2-\Sv t^2[1-A(\dt,t)]}{\sqrt{(z+\lambda)^2-\Sv t^2A(\dt,t)}}\right]\right.
\nonumber\\[0.5mm]
&&~~~~~~~~
+\left.\int\limits_{z^{(-)}}^{\infty}dz\Psi(z,\tau_g)
\left[1-\frac{1}{2}\yFt\sum_{\lambda=\pm 1}\lambda
\frac{(z+\lambda)^2-\Sv t^2[1-A(\dt,t)]}{\sqrt{(z+\lambda)^2-\Sv t^2A(\dt,t)}}\right]\right\},
\nonumber
\end{eqnarray}
\end{adjustwidth}

\noindent
where

\begin{equation}
z^{(-)}=z^{(-)}\zgg=1-\Fv t\sqrt{A(\dt,t)}.
\label{eq25}
\end{equation}

Note that the polarization tensor in the plasmonic region under the inequality (\ref{eq21})
is the complex quantity. The first contributions to (\ref{eq13}) defined in (\ref{eq22}) are
complex due to (\ref{eq23}). As to the second contributions to (\ref{eq13}) defined in
(\ref{eq24}), they are complex due to the square root in the denominators of (\ref{eq24})
which have an imaginary part in the $z$ interval near $z^{(-)}$. Much care must be taken
to this interval in numerical computations.

We are coming now to the region (\ref{eq15}) and rewrite the results of \cite{41,50} for
the polarization tensor in terms of the dimensionless quantities (\ref{eq9}) and (\ref{eq10}).
In this region, the first contributions to  (\ref{eq13}) are

\begin{eqnarray}
&&
\tP_{00}^{(0)}\zgg=\frac{2\alpha}{\sqrt{\Sv t^2-1}}\,\chi\left(\frac{\dt}{\sqrt{\Sv t^2-1}}\right),
\nonumber\\[-1mm]
&&\label{eq26}\\[-1mm]
&&
\tP^{(0)}\zgg={2\alpha}{\sqrt{\Sv t^2-1}}\,\chi\left(\frac{\dt}{\sqrt{\Sv t^2-1}}\right),
\nonumber
\end{eqnarray}

\noindent
where

\begin{equation}
\chi(x)=x+(1-x^2)\,{\rm arctan}\frac{1}{x}.
\label{eq27}
\end{equation}

The second contributions to (\ref{eq13}) in the region (\ref{eq15}) are
expressed as

\begin{adjustwidth}{-\extralength}{0cm}
\begin{eqnarray}
&&
\tP_{00}^{(1)}\ogg=\frac{8\alpha}{\Sv t^2}\int\limits_{\dt}^{\infty}dz\Psi(z,\tau_g)
\left[1+\frac{1}{2}\sum_{\lambda=\pm 1}\lambda
\frac{\Sv t^2-(z+\lambda)^2}{\sqrt{[\Sv t^2-(z+\lambda)^2](\Sv t^2-1)+\Sv t^2\dt^2}}\right],
\nonumber\\[-0.5mm]
&&\label{eq28} \\[0.5mm]
&&
\tP^{(1)}\ogg=\frac{8\alpha}{\Sv t^2}\int\limits_{\dt}^{\infty}dz\Psi(z,\tau_g)
\left[1+\frac{1}{2}\sum_{\lambda=\pm 1}\lambda
\frac{(z+\lambda)^2(\Sv t^2-1)-\Sv t^2\dt^2}{\sqrt{[\Sv t^2-(z+\lambda)^2](\Sv t^2-1)+
\Sv t^2\dt^2}}\right],
\nonumber
\end{eqnarray}
\end{adjustwidth}

{}From (\ref{eq26}) and (\ref{eq28}) it is seen that in the region (\ref{eq15}) the
first contributions to the polarization tensor, $\tP_{00}^{(0)}$ and $\tP^{(0)}$,
are real whereas the second, $\tP_{00}^{(1)}$ and $\tP^{(1)}$, are complex.

In the end of this section, we rewrite the proper nonequilibrium contribution (\ref{eq5})
to the Casimir-Polder force it terms of the dimensionless variables (\ref{eq9}),
(\ref{eq10}), and (\ref{eq12})

\begin{eqnarray}
&&
F_{r}\adt=\frac{\hbar c\alpha(0)}{16\pi a^5}
\int\limits_{0}^{\infty}\!\!dx\,x^4\Theta(\tau_E,\tau_g)\!\!
\int\limits_{1}^{\infty}\!\!t\,dte^{-x\sqrt{t^2-1}}
\nonumber \\
&&~~~~
\times
{\rm Im}\left[(2t^2-1)\Mr(\omega_cx,t,\dt,\tau_g)+\Er(\omega_cx,t,\dt,\tau_g)\right],
\label{eq29}
\end{eqnarray}

\noindent
where
\begin{equation}
\Theta(\tau_E,\tau_g)=\frac{1}{e^{\tau_E}-1}-
\frac{1}{e^{\tau_g}-1}, \qquad
\tau_{E,g}\equiv\tau_{E,g}(\omega_cx)=\frac{\hbar \omega_cx}{k_BT_{E,g}}
\label{eq30}
\end{equation}

\noindent
and the reflection coefficients $R_{\rm TM,TE}^{\rm sub}$ expressed via the dimensionless
variables are contained in (\ref{eq11}).

\section{Equilibrium and Nonequilibrium Casimir-Polder Forces from Fused Silica
Plate Compared to Equilibrium Force in the Presence of Graphene Coating}

In this and in the next sections, we use the formalism presented in Section~2 for
numerical computations of the Casimir-Polder force acting on a nanoparticle
on the source side of the graphene-coated substrate. As the substrate material,
we choose fused silica glass, SiO$_2$, which is in often use both in applications
of graphene in nanoelectronics \cite{53,54,55,56} and in physical experiments
on measuring the Casimir force \cite{57,58}.

For better understanding of the relative roles of substrate, of nonequilibrium effects,
and of graphene coating, in this section we start from calculation of the equilibrium
Casimir-Polder forces acting on a nanoparticles from the uncoated and coated with gapped
graphene SiO$_2$ substrate. Next, the nonequilibrium force from an uncoated SiO$_2$
substrate will be computed.

The equilibrium Casimir-Polder force acing on a nanoparticle from both uncoated and
graphene-coated SiO$_2$ substrates is given by (\ref{eq3}),

\begin{equation}
F_{\rm eq}(a,\Delta,T_E)=F_{M}(a,\Delta,T_E,T_E),
\label{eq31}
\end{equation}

\noindent
where we put $T_E=T_g$.

If the substrate is uncoated, the reflection coefficients $R_{\rm TM,TE}^{\rm sub}$ are
given by (\ref{eq7}) where $\omega=i\xi_{E,l}$ and $\Pi_{00}=\Pi=0$.  If the substrate
is coated with graphene, one should use (\ref{eq7}) with $T_g=T_E$. In this case, the
polarization tensor of graphene is given by expressions (\ref{eq26}) and (\ref{eq28})
found in the frequency region (\ref{eq15}), where, after returning to the dimensional
variables, one can immediately put $\omega=i\xi_{E,l}$ (the explicit expressions for the
polarization tensor of graphene at the pure imaginary frequencies $i\xi_{E,l}$ are
contained in \cite{59}).

\begin{figure}[H]
\vspace*{-8.7cm}
\centerline{\hspace*{-2.7cm}
\includegraphics[width=7.5in]{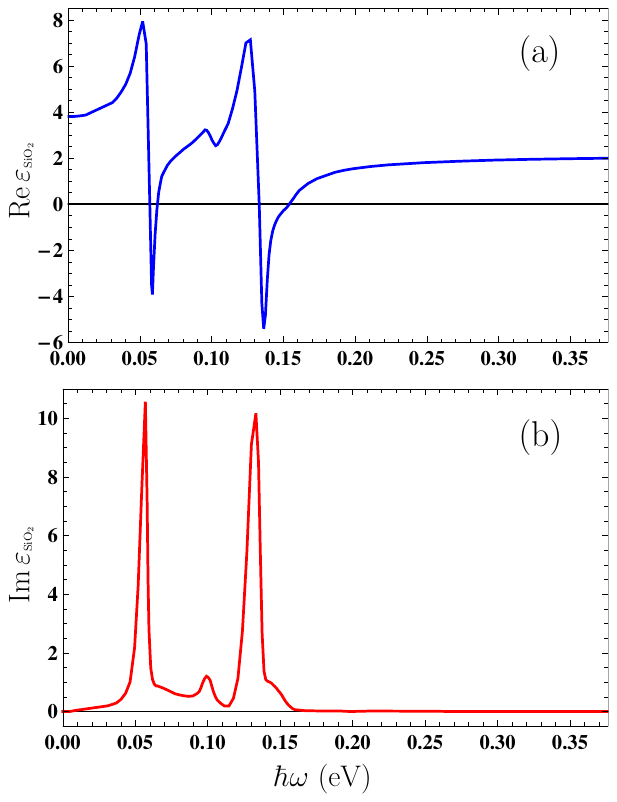}}
\vspace*{-5cm}
\caption{\label{fg1}
Optical data for the (a) real and (b) imaginary parts of the dielectric permittivity of
fused silica glass are shown as the function of frequency along the real frequency axis.}
\end{figure}

Computations of the equilibrium force $F_{\rm eq}$ and the contribution $F_M$ to the
nonequilibrium force require the dielectric permittivity of SiO$_2$ along the imaginary
frequency axis at our disposal, whereas computations of the contribution $F_r$ to the
nonequilibrium force require $\ve_{{\rm SiO}_2}(\omega)$ along the axis of real frequencies.
In Figure~\ref{fg1}, we present the real (a) and imaginary (b) parts of the dielectric
permittivity of SiO$_2$ along the real frequency axis plotted by the tabulated optical data
for the complex index of refraction of SiO$_2$ \cite{60}. The data collected in \cite{60}
extend from 0.0025 to 2000~eV, whereas in Figure~\ref{fg1} the most important region for
our computations is presented. The values of the dielectric permittivity along the
imaginary frequency axis are obtained from ${\rm Im}\,\ve_{{\rm SiO}_2}$ by means of
the Kramers-Kronig relations and can be found in many literature sources (see, e.g.,
\cite{7}).

Now we compute the equilibrium Casimir-Polder force (\ref{eq31}) acting on a nanoparticle
from either an uncoated or graphene-coated SiO$_2$ plate as a function of separation at
the plate temperature $T_p=T_E=300~$K. Keeping in mind that the force values strongly
depend on separation, we normalize them to the Casimir-Polder force on a nanoparticle
from the ideal metal plane at zero temperature \cite{7}

\begin{equation}
F_{CP}^{(0)}(a)=-\frac{3\hbar c}{2\pi a^5}\,\alpha(0).
\label{eq32}
\end{equation}

Taking into account that the Dirac model is applicable at energies below 3~eV \cite{61},
we consider the nanoparticle-plate separations exceeding 200~nm where the characteristic
energy $\hbar\omega_c$ is much less than 3~eV. From the above, we restrict the separation
region by $2~\upmu$m, where the condition (\ref{eq1}) is yet applicable.

\begin{figure}[H]
\vspace*{-10.5cm}
\centerline{\hspace*{-2.7cm}
\includegraphics[width=7in]{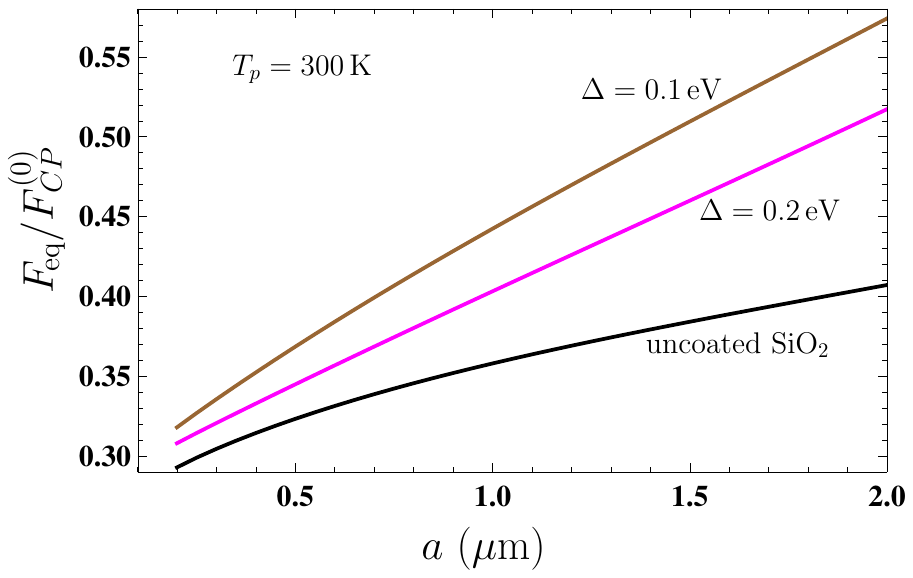}}
\vspace*{-5.1cm}
\caption{\label{fg2}
The ratio of the equilibrium Casimir-Polder force acting on a nanoparticle from
a plate at $T_p=T_E=300~$K to that from an ideal metal plane at  $T_p=T_E=0$ is
shown as the function of separation by the three lines plotted for a uncoated
SiO$_2$ plate and coated by a graphene sheet with the energy gap $\Delta=0.2~$eV
and 0.1~eV.}
\end{figure}

In Figure~\ref{fg2}, the computational results for $F_{\rm eq}/F_{CP}^{(0)}$ at
$T_p=T_E=300~$K are shown as the function of separation by the three lines counted from
bottom to top for an uncoated SiO$_2$ plate and for coated by a graphene sheet with the
energy gap $\Delta=0.2~$eV and 0.1~eV, respectively (in the two latter cases the
temperature of graphene is $T_g=T_p=T_E$). As is seen in Figure~\ref{fg2}, the presence
of graphene coating increases the magnitude of the Casimir-Polder force acting on a
nanoparticle. This increase, however, is larger for a graphene sheet with smaller
energy gap.

To obtain the absolute values of force acting on a nanoparticle one can use the data
of Figure~\ref{fg2} in combination with the following expressions for the static
polarizabilities:

\begin{equation}
\alpha(0)=R^3\frac{\ve(0)-1}{\ve(0)+2}, \qquad \alpha(0)=R^3,
\label{eq33}
\end{equation}

\noindent
which are valid for dielectric and metallic nanoparticles, respectively \cite{26}.
Thus, for SiO$_2$ one obtains $\ve(0)=3.81$ from Figure~\ref{fg1}(a).

Now we calculate the nonequilibrium Casimir-Polder force $F_{\rm neq}^{{\rm SiO}_2}$
acting on a nanoparticle from the uncoated SiO$_2$ plate. In so doing we assume that
the plate temperature is either lower, $T_p=77~$K, or higher, $T_p=500~$K than the
environmental temperature $T_E=300~$K. The computations are performed by Equations
(\ref{eq2}), (\ref{eq3}), and (\ref{eq29}) with the reflection coefficients (\ref{eq7})
where $\Pi_{00}=\Pi=0$ in the absence of graphene coating.

Note that the computation of the proper nonequilibrium contribution $F_r$ to the
nonequlibrium Casimir-Polder force, which is performed along the real frequency axis,
is much more labor and time consuming, especially in the presence of graphene
coating (see in the next section). These computations were realized by using a
program written in the C++ programming language \cite{50} at the Supercomputer
Center of the Peter the Great Saint Petersburg Polytechnic University.

\begin{figure}[H]
\vspace*{-6cm}
\centerline{\hspace*{-2.7cm}
\includegraphics[width=7.5in]{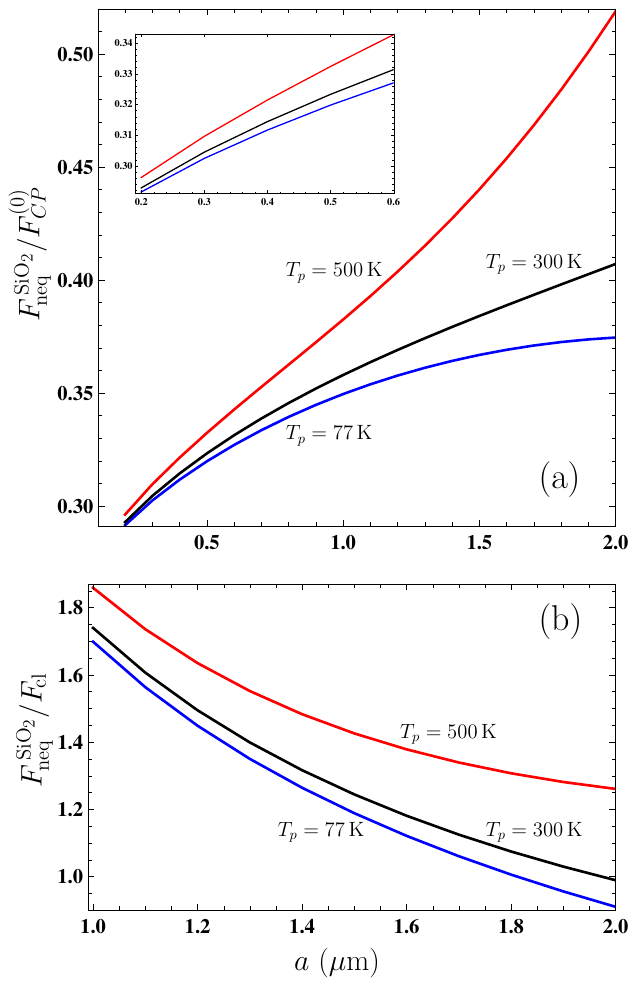}}
\vspace*{-5cm}
\caption{\label{fg3}
The ratio of the nonequilibrium Casimir-Polder force acting on a nanoparticle from
a SiO$_2$ plate  to the equilibrium one from an ideal metal plane (a) at  $T_p=T_E=0$
and (b) at  $T_p=T_E=300~$K  (the classical limit)  is shown as the function
of separation. The bottom and top lines are for the plate temperatures
$T_p=77~$K and 500~K, respectively. The middle lines demonstrate similar ratio when
$T_p=T_E=300~$K. In the inset, the region of short separations
is shown on an enlarged scale.
}
\end{figure}

In Figure~\ref{fg3}(a), we present the computational results for the ratio
$F_{\rm neq}^{{\rm SiO}_2}/F_{CP}^{(0)}$ as the function of separation by the
bottom and top lines for the plate temperatures $T_p=77~$K and 500~K, respectively.
For comparison purposes, the equilibrium Casimir-Polder force ratio
$F_{\rm eq}^{{\rm SiO}_2}/F_{CP}^{(0)}$ for a plate kept at the environmental
temperature $T_p=T_E=300~$K is shown by the middle line which reproduces the bottom
line from Figure~\ref{fg2}. In the inset, the region of small separations
from 0.2 to $0.6~\upmu$m is shown on an enlarged scale for better visualization.

{}From Figure~\ref{fg3}(a), it is seen that the effects of nonequilibrium have a strong
impact on the equilibrium Casimir-Polder force from a SiO$_2$ plate making its magnitude
larger if $T_p>T_E$ and smaller if $T_p<T_E$. From Figure~\ref{fg3}(a) one can also see
that at separations exceeding $1~\upmu$m the Casimir-Polder force (\ref{eq32}) from an
ideal metal plane poorly reproduces the dependence of the nonequilibrium force on the
separation distance.  Because of this, in the region from 1 to $2~\upmu$m, we consider
the same computational results for $F_{\rm neq}^{{\rm SiO}_2}$, but normalize them to
the thermal  Casimir-Polder force from an ideal metal plane at large separations (the
so-called classical limit) \cite{7}

\begin{equation}
F_{\rm\! cl}(a,T_E)=-\frac{3k_BT_E}{4a^4}\,\alpha(0).
\label{eq34}
\end{equation}

In Figure~\ref{fg3}(b), the computational results for the ratio
$F_{\rm neq}^{{\rm SiO}_2}/F_{\rm\! cl}$ are shown as the function of separation by
the bottom and top lines for the plate temperatures $T_p=77~$K and 500~K, respectively.
The middle line demonstrates the  ratio $F_{\rm eq}^{{\rm SiO}_2}/F_{\rm\! cl}$ in
the case of thermal equilibrium $T_p=T_E=300~$K. Figure~\ref{fg3}(b) confirms the
conclusions already made from Figure~\ref{fg3}(a). It is seen also that the effects
of nonequilibrium due to higher temperature than the environmental one make a greater
impact on the equilibrium force than those due to a decrease of temperature to below
the  environmental one.
{{Further decrease of temperature to
below 77~K leads to only an insignificant decrease of the magnitude of
nonequilibrium Casimir-Polder force.}}

\section{Nonequilibrium Casimir-Polder Force from Fused Silica
Plate Coated with Gapped Graphene }

In the previous section, we already considered the nonequilibrium Casimir-Polder force from
a SiO${}_2$ plate. The main difference of the case of a graphene-coated plate considered now
is that the response of graphene to the electromagnetic field described by the polarization
tensor strongly depends on temperature. This is not the case for a SiO${}_2$ plate whose
dielectric permittivity is temperature-independent. As a result, for an uncoated SiO${}_2$
plate the effects of nonequilibrium influence the force only through the factor $\Theta$
in (\ref{eq5}) defined in (\ref{eq6}).

According to (\ref{eq2}), the total nonequilibrium force is the sum of two contributions
$F_M$ and $F_r$. The first of them has the same form as the equilibrium Casimir-Polder
force, whereas the second is the proper nonequilibrium contribution. In order to
understand physics of the process, we consider each of them separately starting
from $F_M$ (for an uncoated SiO${}_2$ plate $F_M$ is equal to the equilibrium force at
$T=300~$K shown by the bottom line in Figure~\ref{fg2} for any temperature of the plate).

\begin{figure}[H]
\vspace*{-8.7cm}
\centerline{\hspace*{-2.7cm}
\includegraphics[width=7.5in]{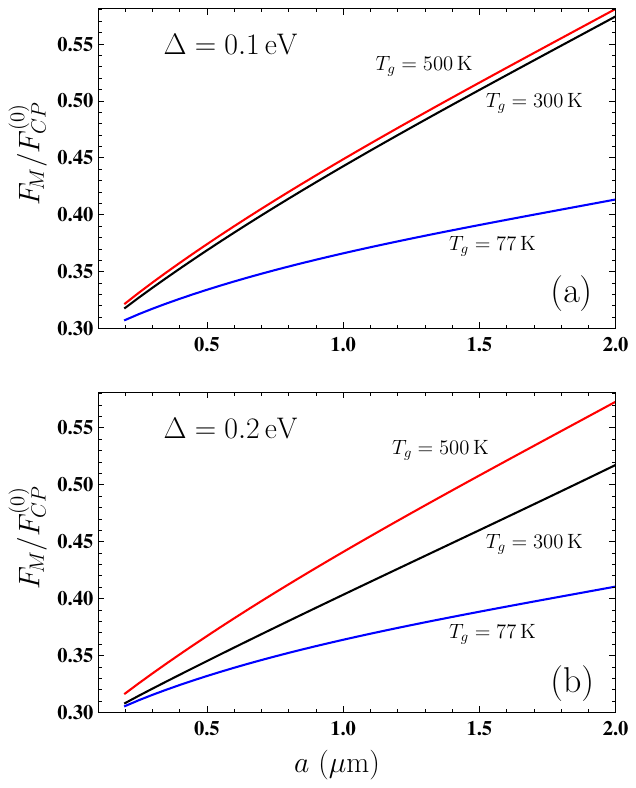}}
\vspace*{-5cm}
\caption{\label{fg4}
The ratio of the first contribution to the nonequilibrium Casimir-Polder force acting
on a nanoparticle from a SiO$_2$ plate coated by a graphene sheet with the energy
gap (a) $\Delta=0.1~$eV and (b) $\Delta=0.2~$eV
to the equilibrium one from an ideal metal plane at  $T_p=T_E=0$
is shown as the function
of separation. The bottom and top lines are for the graphene-plate temperatures
$T_g=77~$K and 500~K, respectively. The middle lines demonstrate similar ratio where
$T_g=T_E=300~$K.
}
\end{figure}

Computations of $F_M$ were performed by (\ref{eq3}), (\ref{eq7}), and the polarization
tensor defined at the pure imaginary frequencies $i\xi_{E,i}$ (see Section~3 and \cite{59}).
The computational results for the ratio $F_M/F_{CP}^{(0)}$ are presented in
Figure~\ref{fg4}(a,b) as the function of separation for (a) graphene coating with the
energy gap $\Delta=0.1~$eV and (b) graphene coating with  $\Delta=0.2~$eV by the bottom
and top lines plotted for the graphene (and plate) temperatures $T_p=77~$K and 500~K.
The middle line shows the ratio $F_{\rm eq}/F_{CP}^{(0)}$ when the graphene temperature
is equal to that of the environment, $T_g=T_E=300~$K.

Unlike the case of an uncoated SiO${}_2$ plate, in the presence of graphene coating
$F_M$ is not equal to the equilibrium force at $T_g=T_E=300~$K. As is seen in
Figure~\ref{fg4}(a), for the relatively small energy gap of the graphene coating
$\Delta=0.1~$eV, the quantity $F_M$ at $T_g=500~$K differs little from the
equilibrium force at $T_g=T_E=300~$K over the entire separation region from 0.2 to
$2~\upmu$m. Thus, the relative deviation

\begin{equation}
\delta F_M\adt=\frac{ F_M\adt- F_{\rm eq}\adt}{ F_{\rm eq}\adt}
\label{eq35}
\end{equation}

\noindent varies in this case from 1.21\% at $a=0.2~\upmu$m to 1.14\% at $a=2~\upmu$m,
i.e., is almost independent on separation.

{}From Figure~\ref{fg4}(a) it is also seen that both the middle and top lines grow
with separation. This means that for a substrate coated by graphene with
$\Delta=0.1~$eV considerable contribution to both $F_M$ at $T_g=500~$K and
$F_{\rm eq}$ at $T_g=T_E=300~$K is given by the term of the Lifshitz formula
with $l=0$. At the same time, for a graphene coating with $\Delta=0.1~$eV
at $T_g=77~$K the deviation of $F_M$ from $F_{\rm eq}$ is much larger and it
increases with increasing separation [see the bottom and middle lines in
Figure~\ref{fg4}(a)]. Thus, at $a=0.2~\upmu$m we have $\delta F_M=-3.35$\% and
at $a=2~\upmu$m already $\delta F_M=-28$\%

The comparison of Figure~\ref{fg4}(a) plotted for a graphene coating with $\Delta=0.1~$eV
with Figure~\ref{fg4}(b) plotted for $\Delta=0.1~$eV shows that the values of $F_M$
computed at $T_g=77~$K almost coincide (the difference of 0.6\% at $a=0.2~\upmu$m and
1\% at $a=2~\upmu$m). This means that at the relatively low $T_g=77~$K the thermal
corrections are rather small and depend only slightly on the value of $\Delta$ in the
range of separations from 0.2 to $2~\upmu$m. The comparison of the values of $F_M$ at
$T_g=500~$K for graphene coatings with $\Delta=0.1$ and 0.2~eV [top lines in
Figures~\ref{fg4}(a) and \ref{fg4}(b)] also demonstrates rather small deviations
(1\% at $a=0.2~\upmu$m and 1.9\% at $a=2~\upmu$m). This is because at so high temperature
the thermal corrections are rather large and have only a weak dependence on $\Delta$.
As to the equilibrium Casimir-Polder force at $T_g=T_E=300~$K, it depends on $\Delta$
more distinctly (compare with Figure~\ref{fg2}). Note also that the magnitudes of $F_M$
at both 77~K and 500~K are larger than the magnitudes of $F_{\rm neq}$ from an
uncoated SiO$_2$ plate at the same respective temperatures. This is seen from the
comparison of Figure~\ref{fg4}(a,b) and Figure~\ref{fg3}.

\begin{figure}[H]
\vspace*{-9.cm}
\centerline{\hspace*{-2.7cm}
\includegraphics[width=7.5in]{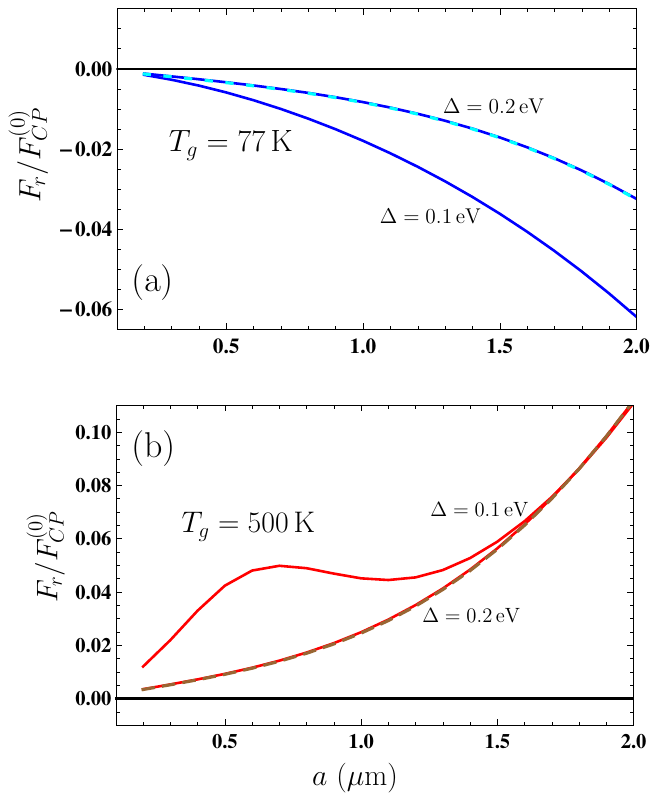}}
\vspace*{-5.cm}
\caption{\label{fg5}
The ratio of the second contribution to the nonequilibrium Casimir-Polder force acting
on a nanoparticle from a SiO$_2$ plate coated by a graphene sheet (a) at $T_g=77~$K
and (b) $T_g=500~$K to the equilibrium one from an ideal metal plane at  $T_p=T_E=0$
is shown as the function of separation (a) by the two lines for $\Delta=0.1$ and 0.2~eV,
where the latter coincides with that for an uncoated plate shown by the dashed line,
 and (b) by the two solid
lines for $\Delta=0.1$ and 0.2~eV,
where the latter coincides with that for an uncoated plate shown by the dashed line 
}
\end{figure}

Next, we consider the second contribution, $F_r$, to the nonequilibrium Casimir-Polder
force (\ref{eq2}) acting on a nanoparticle from a graphene-coated substrate.
The numerical computations of $F_r$ were performed in the dimensionless variables
by (\ref{eq11}), (\ref{eq29}) using the polarization tensor in (\ref{eq17}), (\ref{eq19}),
(\ref{eq22}), (\ref{eq24}), (\ref{eq26}), and (\ref{eq28}). The computational results
for the ratio $F_r/F_{CP}^{(0)}$ are presented  in Figure~\ref{fg5} as the function of
separation for (a) graphene-plate temperature $T_g=77~$K and (b) graphene-plate temperature
$T_g=500~$K. The solid line in Figure~\ref{fg5}(a) is plotted for a graphene coating with
$\Delta=0.1~$eV, whereas another (dashed) line presents the coinciding results for
the graphene coating with $\Delta=0.2~$eV and for the uncoated SiO$_2$ plate.
The top and bottom solid lines in Figure~\ref{fg5}(b) are for a graphene coating with
$\Delta=0.1$ and 0.2~eV, respectively, and the dashed line is for the uncoated SiO$_2$
plate. The form of presentation in Figure~\ref{fg5}(a,b) provides a way to determine the
comparative contributions of the regions (\ref{eq14}) and (\ref{eq15}) to $F_r$.

As is seen in Figure~\ref{fg5}(a,b), the quantity $F_r$, unlike $F_M$, substantially
depends on the value of $\Delta$ at both $T_g=77$ and 500~K. The point is that for
$T_g<T_E$ the contribution to $F_r$ from the regions (\ref{eq14}) and (\ref{eq15}) 
are positive, i.e.,  decrease the force
magnitude. The opposite situation occurs for $T_g>T_E$,
i.e., the contribution of regions (\ref{eq14}) and (\ref{eq15}) to $F_r$ 
are negative leading to the increase
of force magnitude.

For the nonequilibrium Casimir-Poder force acting on a nanoparticle from a freestanding
in vacuum graphene sheet, different contributions to $F_r$ were investigated in \cite{50}.
It was shown that for a graphene with $\Delta=0.2~$eV at $T_g=77~$K it holds $F_r\approx 0$,
whereas for $\Delta=0.1~$eV the main contribution to $F_r$ is given by the region (\ref{eq15})
which leads to the increase of the force magnitude.  If the graphene with $\Delta=0.1~$eV
is heated up to $T_g=500~$K, at short separations ($a\lesssim 0.4~\upmu$m) the contribution
from the region (\ref{eq15}) is dominant. With increasing separation, however, the
region (\ref{eq14}) takes the main role leading to a minor increase of force magnitude.
For a freestanding graphene sheet with $\Delta=0.2~$eV, the main contribution to $F_r$
at all separations is given by the region (\ref{eq15}) \cite{50}.

For a graphene-coated substrate one has a more complicated situation. The point is that
for an uncoated SiO$_2$ plate it is just $F_r$ which determines a distinction of the
nonequilibrium force from the equilibrium one. As s result, for an uncoated plate $F_r>0$
for $T_g<T_E$ and $F_r<0$ for $T_g>T_E$. As is seen in Figure~\ref{fg5}(a), for
$T_g=77~\mbox{K}<T_E$ the graphene coating with $\Delta=0.2~$eV does not influence
the value of $F_r$, which is fully determined by the properties of a substrate.
However, the graphene coating with $\Delta=0.1~$eV makes a sizable effect on $F_r$.

\begin{figure}[H]
\vspace*{-9.cm}
\centerline{\hspace*{-2.7cm}
\includegraphics[width=7in]{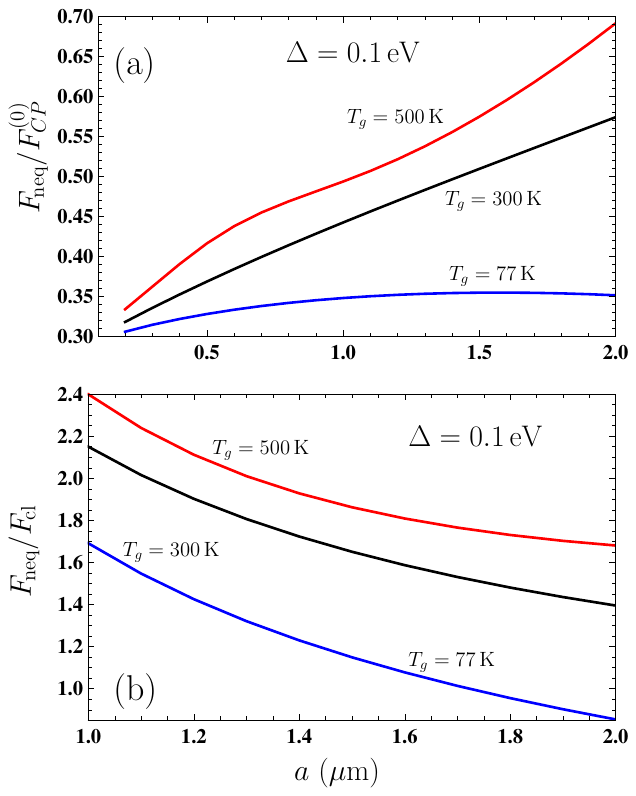}}
\vspace*{-5.2cm}
\caption{\label{fg6}
The ratio of the nonequilibrium Casimir-Polder force acting on a nanoparticle from
a SiO$_2$ plate coated by a graphene sheet with $\Delta=0.1~$eV
to the equilibrium one from an ideal metal plane (a) at  $T_p=T_E=0$
and (b) at  $T_p=T_E=300~$K  (the classical limit)  is shown as the function
of separation. The bottom and top lines are for the graphene-plate temperatures
$T_g=77~$K and 500~K, respectively. The middle lines demonstrate similar ratio where
$T_g=T_E=300~$K.
}
\end{figure}

\begin{figure}[H]
\vspace*{-6cm}
\centerline{\hspace*{-2.7cm}
\includegraphics[width=7.5in]{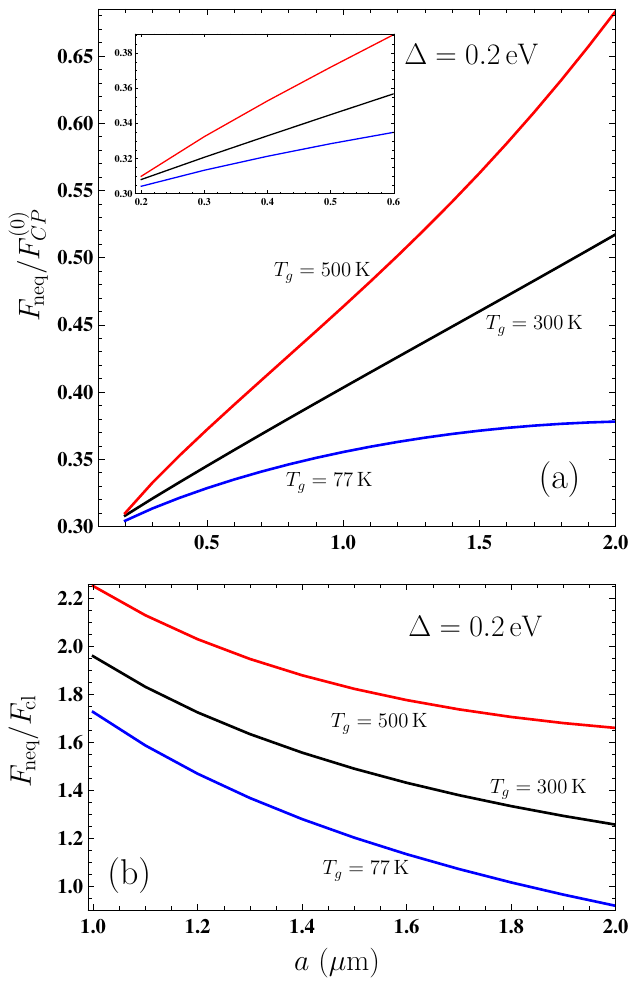}}
\vspace*{-5cm}
\caption{\label{fg7}
The ratio of the nonequilibrium Casimir-Polder force acting on a nanoparticle from
a SiO$_2$ plate coated by a graphene sheet with $\Delta=0.2~$eV
to the equilibrium one from an ideal metal plane (a) at  $T_p=T_E=0$
and (b) at  $T_p=T_E=300~$K  (the classical limit)  is shown as the function
of separation. The bottom and top lines are for the graphene-plate temperatures
$T_g=77~$K and 500~K, respectively. The middle lines demonstrate similar ratio where
$T_g=T_E=300~$K. In the inset, the region of short separations
is shown on an enlarged scale.
}
\end{figure}

Now we are in a position to present the computational results for the total
nonequilibrium Casimir-Polder force $F_{\rm neq}$ between a nanoparticle and
a graphene-coated substrate which is the sum of contributions $F_M$ given  in
Figure~\ref{fg4} and $F_r$ given in Figure~\ref{fg5}. In Figures~\ref{fg6}(a,b) and
\ref{fg7}(a,b) the results for $F_{\rm neq}$ are shown for the graphene coating with
$\Delta=0.1$ and 0.2~eV, respectively. 

In each figure, the bottom, middle, and top lines
are plotted for the graphene-plate temperature $T_g=77~$K, 300~K, and 500~K,
respectively. In Figures~\ref{fg6}(a) and \ref{fg7}(a), the values of $F_{\rm neq}$
are normalized to the zero-temperature Casimir-Polder force $F_{CP}^{(0)}$ from an
ideal metal plane (\ref{eq32}), whereas in  Figures~\ref{fg6}(b) and \ref{fg7}(b)
--- to the classical limit of the Casimir-Polder force $F_{\rm cl}$ from an
ideal metal plane (\ref{eq34}).

Thus, the middle lines in Figures~\ref{fg6}(a) and \ref{fg7}(a) reproduce the top
and middle lines in Figure~\ref{fg2}, respectively, which are also plotted for the
equilibrium Casimir-Polder force from the graphene-coated substrate with $\Delta=0.1$
and 0.2~eV. Figures~\ref{fg6}(b) and \ref{fg7}(b) can be compared with Figures~1 and 2
in \cite{50}, plotted there for a freestanding in vacuum graphene sheet, in order to
determine an impact of the SiO$_2$ substrate on the nonequilibrium force.
By and large, Figures~\ref{fg6} and \ref{fg7} are in analogy to Figure~\ref{fg3}
related to the case of an uncoated substrate.

{}From Figures~\ref{fg6} and \ref{fg7} it is seen that the nonequilibrium force
$F_{\rm neq}$ from the
graphene-coated SiO$_2$ plate at both $T_g=77~$K and 500~K is larger in magnitude than
the force $F_{\rm neq}^{{\rm SiO}_2}$ from an uncoated SiO$_2$ plate and from a
freestanding graphene sheet. Thus, for a SiO$_2$ plate coated with a graphene sheet
with $\Delta=0.1~$eV at $T_g=77~$K the ratio $F_{\rm neq}/F_{\rm neq}^{{\rm SiO}_2}$
is equal to 1.05 and 0.94 at separations $a=0.2$ and $2~\upmu$m, respectively.
The same ratio for the same graphene coating but at $T_g=500~$K is equal to 1.12
and 1.33 at the same respective separations. For a graphene coating with
$\Delta=0.2~$eV at $T_g=77~$K, the ratio $F_{\rm neq}/F_{\rm neq}^{{\rm SiO}_2}$
is equal to 1.04 and 1.01 at $a=0.2$ and $2~\upmu$m, and, if the graphene-coated plate
is at the temperature $T_g=500~$K, --- to 1.04 and 1.32, respectively.

If we compare the Casimir-Polder force from the graphene-coated SiO$_2$ plate with
that from a freestanding graphene sheet, $F^{\rm free}$, it is seen that the presence
of a substrate significantly increase the magnitudes of both equilibrium and
nonequilibrium forces. This increase is the most pronounced at short separations.
As an example, for a graphene coating on a SiO$_2$ plate and
a freestanding graphene sheet with $\Delta=0.1~$eV, we obtain
$F_{\rm eq}/F_{\rm eq}^{\rm free}=4.3$ and 1.4 at $a=0.2$ and $2~\upmu$m, respectively.
For the nonequilibrium forces with the same value of $\Delta$,  the ratio
$F_{\rm neq}/F_{\rm neq}^{\rm free}$ is equal to 8.2 and 2.2 at $T_g=77~$K and
to 4.7 and 1.6 at $T_g=500~$K at the same respective separations.

The above results allow to control the value of the nonequilibrium Casimir-Polder
force acting on nanoparticles from the graphene-coated substrates.

\section{Discussion}
The out-of-thermal-equilibrium Casimir and Casimir-Polder forces is a rather
novel subject which is investigated only during the last 25 years. Despite
of this, considerable advances have already been made in understanding the
underlying physics and mathematical description of the effects of
nonequilibrium. Specifically, as noted in Section 1, the fundamental
Lifshitz theory of the Casimir and Casimir-Polder forces was generalized for
the out-of-thermal-equilibrium conditions \cite{19,20,21,22,23,24,25}.
At first this was made for the case when the material properties are
temperature-independent, but in succeeding years generalized for materials
whose response functions to the electromagnetic field explicitly depend on
temperature as a parameter \cite{29,30}.

Graphene is a unique novel material whose response functions described by
the polarization tensor essentially depend on the temperature. Because of
this, the effects of nonequilibrium in the Casimir-Polder force acting on
nanoparticles from a graphene sheet are best suited to both the theoretical
study and practical utility in nanoelectronics. Previously these effects
were considered only for a freestanding graphene sheets \cite{49,50}, which
is a configuration not easily accessible in a laboratory. In this article,
the nonequilibrium Casimir-Polder force on a nanoparticle from a substrate
coated with gapped graphene sheet was investigated as the function of
separation, temperature, and energy gap in the case of the most often
used fused silica glass substrate. This opens a way to a practical
implementation of this physical phenomenon
{{in the field effect transistors and other devices
of bioelectronics \cite{62,63,64}. The obtained results can be also
generalized to other two-dimensional materials and van der Waals
heterostructures  employing different 2D crystals \cite{65,66,67}.}}

\section{Conclusions}

To conclude, in this article the formalism of the Lifshitz theory generalized
to out-of-thermal-equilibrium conditions with temperature-dependent material
properties was used to investigate the nonequilibrium Casimir-Polder force
between nanoparticles and either cooled or heated fused silica plate
substrate coated with gapped graphene sheet. The response of graphene to the
electromagnetic field was described by the polarization tensor in the
framework of the Dirac model.

We investigated two different contributions to the nonequilibrium
Casimir-Polder force and determined their physical meaning and relative
role depending on the frequency and separation regions. The total
nonequilibrium force from a graphene-coated fused silica glass substrate
was compared with the equilibrium one from the same source, as well as
with both equilibrium and nonequilibrium forces from an uncoated silica
glass plate. A comparison with the nonequilibrium Casimir-Polder force
from a freestanding in vacuum graphene sheet has also been made.

It was shown that the nonequilibrium force from the graphene-coated
silica glass substrate kept at both lower and higher temperature than in
the environment is larger in magnitude than the nonequilibrium force
from an uncoated silica glass plate and from a freestanding graphene
sheet. According to the results obtained, an increase of the energy gap
of graphene coating leads to smaller force magnitudes and to a lesser
impact of the graphene coating on the nonequilibrium force acting on a
nanoparticle on the source side of an uncoated silica glass plate.
{{By and large, we determined the impact
of temperature of a graphene sheet, the role of a
substrate and of the nonzero energy gap of graphene coating on the
nonequilibrium Casimir-Polder force between a nanoparticle and a
graphene-coated substrate spaced at different separations.}}

The above results may find application in the rapidly progressing areas
of nanotechnology dealing with integrated nanoparticle-biomolecular
systems.

\vspace{6pt}


\funding{The work of O.Yu.T. was supported by the Russian Science Foundation
under Grant No. 21-72-20029.  G.L.K. was partially funded by the
Ministry of Science and Higher Education of Russian Federation
("The World-Class Research Center: Advanced Digital Technologies,"
contract No. 075-15-2022-311 dated April 20, 2022). The research
of V.M.M. was partially carried out in accordance with the Strategic
Academic Leadership Program "Priority 2030" of the Kazan Federal
University. }

\begin{adjustwidth}{-\extralength}{0cm}

\reftitle{References}


\end{adjustwidth}
\end{document}